\documentclass[epsfig]{mn2e}
\usepackage{epsfig}

\input{epsf}

\twocolumn 

\newcommand{\beq}{\begin{equation}}
\newcommand{\eeq}{\end{equation}}
\newcommand{\beqa}{\begin{eqnarray}} 
\newcommand{\eeqa}{\end{eqnarray}}
\newcommand{\bea}{\begin{array}} 
\newcommand{\ea}{\end{array}}

\font \bolditalics = cmmib10

\font \matrix = cmssdc10
\newcommand{\Vec}[1]{{\textfont1=\bolditalics \hbox{$#1$}}}
\newcommand{\Mat}[1]{{\textfont1=\matrix \hbox{$#1$}}}
\newcommand{\tp}{{\rm t}}

\newcommand{\average}[1]{\left\langle #1 \right\rangle}

\newcommand{\vp}{\varphi}
\newcommand{\vt}{\vartheta}
\newcommand{\ve}{\varepsilon}

\newcommand{\Cov}{\Mat C}

\newcommand{\dd}{{\mathrm d}}

\newcommand{\ontop}[2]{
  \renewcommand{\arraystretch}{0.2}
  \begin{array}{c}
  #1 \\ #2
  \end{array}
  \renewcommand{\arraystretch}{1.0}
}
\newcommand{\lsim}{\ontop{<}{\sim}}

\newcommand{\del}{\partial}

\newcommand{\Omegam}{\Omega_{\rm m}}

\newcommand{\ii}{{\rm i}}
\newcommand{\ee}{{\rm e}}
\newcommand{\gammat}{\gamma_{\rm t}}
\newcommand{\J}{{\rm J}}

\newcommand{\Mapsq}{M_{\rm ap}^2}

\newcommand{\EDIT}{}

\title[Generalised Eigenmode Analysis of Weak Lensing Surveys]
{Designing Weak Lensing Surveys: A Generalised Eigenmode Analysis}
\author[Kilbinger \& Munshi]
{Martin Kilbinger$^{1}$ and Dipak Munshi$^{2,3}$\\
$^{1}$Institut f. Astrophysik u. Extraterrestrische Forschung, Universit{\"a}t Bonn,
Auf dem H{\"u}gel 71, D-53121 Bonn, Germany \\
$^{2}$Institute of Astronomy, Madingley Road,
Cambridge, CB3 OHA, United Kingdom\\
$^{3}$Astrophysics Group, Cavendish Laboratory, Madingley Road, 
Cambridge CB3 OHE, United Kingdom\\
}

\begin{document}
\maketitle

\begin{abstract}
We study the estimators of various second-order weak lensing
statistics such as the shear correlation functions $\xi_{\pm}$ and the
aperture mass dispersion $\langle\Mapsq\rangle$ which can directly be
constructed from weak lensing shear maps. We compare the efficiency
with which these estimators can be used to constrain cosmological
parameters. To this end we introduce the Karhunen-Lo\`eve (KL)
eigenmode analysis techniques for weak lensing surveys.  These tools
are shown to be very effective as a diagnostics for optimising survey
strategies. The usefulness of these tools to study the effect of
angular binning, the depth and width of the survey and noise
contributions due to intrinsic ellipticities and number density of
source galaxies on the estimation of cosmological parameters is
demonstrated.
Results from independent analysis of various parameters and joint
estimations are compared. We also study how degeneracies among various
cosmological and survey parameters affect the eigenmodes associated
with these parameters.
\end{abstract}

\begin{keywords}
Cosmology: theory -- gravitational lensing -- large-scale structure 
of Universe -- Methods: analytical, statistical, numerical
\end{keywords}

\section{Introduction}

Weak gravitational lensing studies (for a review see e.g.\ Mellier
1999; Bartelmann \& Schneider 2001; R\'efr\'egier 2003; van Waerbeke
\& Mellier 2003) probe the correlation of the observed ellipticities
of background galaxies. Cosmic shear provides unbiased estimates of
the matter power spectrum and hence can help us to place independent
constraints on cosmological parameters (see e.g. Hu 1999). The
parameter degeneracies from weak lensing surveys are very different
from those present in CMB studies (van Waerbeke et al. 2002). This
means that even most precise measurements from CMB observations can be
improved by joint analyses with weak lensing data (Contaldi et al.\
2003; Hu \& Tegmark 1999).  Thus, the recent detection of weak lensing
by the large-scale structure (Bacon et al. 2000; Kaiser, Wilson \&
Luppino 2000; Wittman et al.\ 2000.; van Waerbeke et al.\ 2000, 2002;
Maoli et al.\ 2001; R\'efr\'egier, Rhodes \& Growth 2002) has opened a
completely new window to the Universe. While first generations of
cosmic shear surveys have demonstrated the feasibility of weak lensing
studies in constraining the dark matter power spectrum parametrised by
$\Omegam$, $\sigma_8$ and the shape parameter $\Gamma$ (R\'efr\'egier
2003; van Waerbeke, Mellier \& Hoekstra 2005), future surveys will be
able to probe much larger scales and therefore allow us to study the
linear regime directly and put more stringent bounds on cosmological
parameters such as the equation of state of dark energy and its time
variation.

As survey sizes grow the amount of data that is to be
processed also increases.
It is then essential to study data compression techniques which can
effectively compress the data vector to a lower-dimensional space.
The effect of such a compression on the Fisher information matrix can
be studied by analysing the Karhunen-Lo\`{e}ve (KL) eigenmodes
(Karhunen 1947; Lo\`eve 1948) associated with the Fisher matrix. These
techniques are already in use in cosmic microwave background studies
(see e.g. Bond 1995 for COBE \& FIRAS data, Bunn 1995 for COBE data
and Tegmark, Taylor \& Heavens 1997, hereafter TTH, for a general
review of the KL eigenmode analysis). In the context of galaxy
surveys, Matsubara, Szalay \& Landy (2000) applied KL methods to the
Las Campanas redshift survey, Hamilton, Tegmark \& Padmanabhan (2000)
studied PSCz using KL techniques and Szalay et al.~(2003) have used
this method to study data from SDSS.  Watkins et al.~(2001) have used
these techniques to analyse the peculiar velocity surveys for removal
of small-scale, non-linear modes from large-scale linear modes which
retain cosmological information. Other studies using the KL eigenmode
methods for velocity field surveys include Hoffmann \& Zaroubi (2000)
and Silberman et al.~(2001).  Recent studies by Huterer \& White
(2005) have underlined its importance for filtering unwanted
non-cosmological information at small angular scales.

The data vector in weak lensing surveys which is used to extract
cosmological information is usually the shear correlation at various
angular separations, estimated from the shapes of the observed
background galaxies. Thus, it represents an already compressed data
set for which likelihood analyses are feasible although further
compression can speed up the analysis. Moreover, for
third-order statistics, the number of measured triangle configurations
can be very large even in a binned way and and further compression of
the data might be necessary to place constraints in a
high-dimensional cosmological parameter space.  Therefore, the study of KL methods
in the context of cosmic shear will be very useful for future weak
lensing surveys.

Being not only an effective tool for data compression, the KL
eigenmode analysis can also help to understand the specific linear
combination of angular scales and the redshift range that contributes
most significantly to a specific estimator. A KL eigenmode analysis of
weak lensing observables in redshift space has already been done in by
Heavens (2003), where the distribution of observed ellipticities was
directly used as data vector. We extend these analyses to second-order
shear statistics as functions of projected angular scales and for
realistic survey geometries. The latter issue is particularly
interesting as sky coverage in any weak lensing survey will always be
non-trivial due to the presence of bright foreground stars and
galaxies. Our study therefore gives us useful clues about the survey
design. We provide a KL eigenmode analysis for the second-order shear
measures $\xi_{\pm}$, $\xi_{\rm tot}$ and $\langle\Mapsq\rangle$ (to
be defined later). The covariance of these estimators were studied in
detail by Schneider et al.~(2002) to assess the effect of source
ellipticity dispersion and finite sky coverage on the estimation of
cosmological parameters. The influence of survey geometry was
investigated in Kilbinger \& Schneider (2004). We use those results to
construct the Fisher matrix to forecast parameter constraints. In
contrast to those previous studies, where the above mentioned
second-order shear statistics were used to determine cosmological
parameters, we also consider the parameter dependence of the
covariance of these estimators and its influence on the Fisher matrix.
Moreover, the KL approach allows us to study in detail how the
convergence power spectrum is sampled on different scales by a given
shear survey.


The paper is arranged in the following way. Section \S
\ref{sec:notation} is devoted to a recapitulation of the basics of the
KL eigenmode analysis, reformatted in the weak lensing context.  In
section \S 3, we describe our simulations and survey strategies in
detail (\S 3.1 -- \S 3.3) and present the results for the eigenmode
analysis (\S 3.4), including the effect of survey
strategy, noise and binning. We discuss our findings in \S 4.

\section{Notations and Formalism}
\label{sec:notation}

\subsection{Survey characteristics}
\label{sec:survey_char}

We consider several different survey strategies in this work. First,
individual images are distributed randomly (but non-overlapping)
within circular patches on the sky, where `image' means one individual
field of view of size $13^\prime \times 13^\prime$. The individual
patches are uncorrelated, thus the largest scale where the shear
correlation can be measured is twice the patch radius. The random
image placement accounts for the fact that for realistic
surveys, bright stars and foreground galaxies should be avoided and
therefore sparse sampling of a sky region is necessary.

A survey consists of $P$ patches of radius $R$, each patch containing
$N$ images. The total number of images is 300 for all patch
geometries, corresponding to 14.1 square degrees for the entire
survey.

We use $N=30$ and $R=100^\prime, 140^\prime$,
respectively. Note that the number of patches in a survey is $P =
300/N$.  We denote these patch geometries by the two numbers $(N,R)$,
thus we have the configurations $(30,100^\prime)$ and
$(30,140^\prime)$.

\begin{figure}
\protect\centerline{
\epsfysize = 2. truein
\epsfbox[158 367 576 664]
{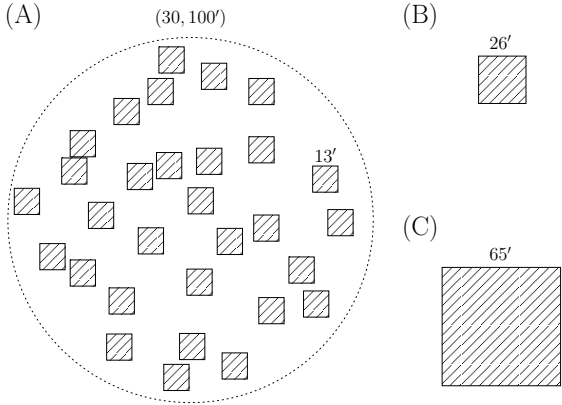}}
\caption{Sketch of the two kinds of survey strategies used in this
  work. On the left, a patch of $100\prime$ radius is filled
  sparsely with 30 images of size $13^\prime$, a survey consists of
  10 such patches. For comparison, on the right the field sizes of
  surveys consisting of 75 uncorrelated $26^\prime \times
  26^\prime$ images and 12 independent $65^\prime \times 65^\prime$ 
  fields, respectively, are shown.}
\label{fig:sketch}
\end{figure}

Second, we define two surveys which consist of single,
uncorrelated fields of view with the same total area of 14.1 square
degrees. The first survey consists of 75 uncorrelated $26^\prime \times
26^\prime$ square images -- typical fields of view for current
wide-field cameras like WFI on the ESO 2.2-m telescope. The second survey
represents 12 independent $65^\prime \times 65^\prime$-fields,
corresponding to new-generation wide-field cameras like MegaCam/CFHT.
We denote these surveys with $75 \cdot 26^{\prime \, 2}$
and $12 \cdot 65^{\prime \, 2}$, respectively. A sketch of the
geometry of the surveys used in this work can be found in
Fig.~\ref{fig:sketch}. 

If not indicated otherwise, the number density of source galaxies is
$n_{\rm gal} = 30 \, {\rm arcmin}^{-2}$. This number density of
high-redshift galaxies which are usable for weak lensing shape
measurements can be achieved with high-quality ground-based imaging
data from a 4 m-class telescope. The source galaxy ellipticity
dispersion is $\sigma_\ve = 0.3$, if not stated otherwise.

\subsection{Second-order shear statistics}
\label{sec:soss}

The light of distant galaxies is deflected while propagating through
the tidal gravitational potential of intervening matter
inhomogeneities which constitute the large-scale structure in the
Universe. The image of a galaxy gets distorted, and in the weak
lensing approximation, the relation $\ve = \ve^{\rm s} + \gamma$ holds
between the observed ellipticity $\ve$ of a galaxy, its intrinsic
ellipticity $\ve^{\rm s}$ and the shear $\gamma = \gamma_1 + {\rm i}
\gamma_2$. From the shear, one can in principle reconstruct the projected
matter density or convergence $\kappa$.

For two points separated by a vector $\Vec \theta$ with polar angle $\vp$,
one defines the tangential and cross-component of the shear as
\begin{equation}
  \gamma_{\rm t} \equiv - \, \Re \left( \gamma \ee^{-2 \ii \vp} \right)
  \quad\quad \mbox{and} \quad \quad \gamma_\times \equiv - \, \Im \left(
  \gamma \ee^{-2 \ii \vp} \right).
  \label{gamma-tx-def}
\end{equation}
The two shear two-point correlation functions (2PCF) and their relation to the
convergence power spectrum $P_\kappa$ are (Kaiser 1992)
\begin{equation}
\xi_{\pm}(\theta) = \average{\gammat\gammat} \pm
\average{\gamma_\times\gamma_\times}
= \frac{1}{2 \pi} \int\limits_0^\infty \dd \ell
\, \ell \, P_\kappa(\ell) \J_{0,4}(\ell \theta),
\label{xi-pm-def}
\end{equation}
where the first-kind Bessel function $\J_0$ ($\J_4$) corresponds to the `$+$'
(`$-$') correlation function.

A very useful second-order statistics is the dispersion of the
so-called aperture mass (Kaiser 1995, Schneider 1996), which is the
weighted tangential shear integrated over a circular aperture. When
using a polynomial compensated filter function as weight
(Schneider 1996, Schneider et al.~1998), the dispersion of the
aperture mass is related to the power spectrum by
\begin{equation}
\average{ \Mapsq(\theta)} = \frac{1}{2\pi} \int\limits_0^\infty \dd
\ell \, \ell \, P_\kappa(\ell) \left( \frac{24 \, \J_4(\ell \theta)}{(
\ell \theta )^2} \right)^2.
\label{Map-disp}
\end{equation}
The aperture mass statistics is a measure of the E-mode only of cosmic
shear. A similar statistics $\langle M_\perp^2\rangle$ can be defined
as the weighted cross-component of shear in an aperture, which is a
measure of the B-mode only. The aperture mass statistics yields the
separate measurement of the E- and B-mode which allows one to quantify
contaminations to the gravitational shear signal. A B-mode mainly
comes from systematic measurement errors like imperfect PSF correction
and/or intrinsic alignment of source galaxies, whose shape
correlations are to be measured.

The 2PCF are easily obtained from real data however complicated the
survey geometry might be. The two aperture statistics can be
expressed in terms of the 2PCF:
%
\begin{equation}
\average{M_{{\rm ap}, \perp}^2(\theta)} =
\int\limits_0^{2 \theta} \frac{\dd \vt \, \vt}{2\theta^2} \left[
\xi_+(\vt) T_+\left( \frac{\vt}{\theta} \right) \pm 
\xi_-(\vt) T_-\left( \frac{\vt}{\theta} \right) \right],
\label{Map-xi-EB}
\end{equation}
%
where $T_+$ and $T_-$ are given explicitly in Schneider, van Waerbeke
\& Mellier (2002).  In the absence of a B-mode, both terms on the
right-hand side of the previous equation are of the same amplitude,
resulting in $\langle M_\perp^2\rangle = 0$.

An unbiased estimator of the shear 2PCF (\ref{xi-pm-def}) for a set of
(logarithmic) angular bins is given by
\begin{equation}
\hat \xi_{\pm}(\vt) = \frac{1}{N_{\rm p}(\vt)} \sum_{ij} ( \ve_{i {\rm
    t}} \ve_{j {\rm t}} \pm 
\ve_{i\times} \ve_{j\times} ),
\label{xi-hat}
\end{equation}
where the sum goes over all pairs of galaxies, whose angular
separation is in the angular bin corresponding to $\vt$. $\ve_i$ is
the (complex) ellipticity of a galaxy at position $\vec \theta_i$ and
$N_{\rm p}(\vt)$ is the number of pairs in the $\vt$-bin. Galaxies
measured from real data might carry weight factors reflecting the individual
noise and precision in the shape measurement. Using these weights, one
can define a more optimal estimator for the 2PCF; however, we do not
include this effect and assume equal weights for each galaxy.

Unbiased estimators for the dispersions of the two aperture mass
statistics $\average{\Mapsq}$ and $\average{M_\perp^2}$ (\ref{Map-xi-EB}), respectively,
are
\begin{eqnarray}
  \lefteqn{{\cal M}_\pm(\theta) = \sum_{i=1}^{I} \frac{\Delta \vt_i \vt_i}{2
    \theta^2} \left[ \hat \xi_+(\vt_i) T_+\left( \frac{\vt_i}{\theta}
    \right) \right.}  \nonumber \\
    && \left. \pm \hat \xi_-(\vt_i) T_-\left( \frac{\vt_i}{\theta} \right)
    \right],
  \label{Map-est}
\end{eqnarray}
where the limit $I$ of the sum must be chosen such that $\theta$ is
half the upper limit of the $I$-th bin. For simplicity of notation, we
will from now on identify the estimators with their corresponding
shear statistics, and write $\xi_+, \xi_-$ and $\langle \Mapsq
\rangle$ for the estimators.

These second-order shear estimators are used extensively in the
literature, including independent studies involving $\xi_{\pm}$ or the
joint data vector $\xi_{\rm tot} = (\xi_{+}, \xi_{-})$. We do not
take into account the shear dispersion in an aperture $\langle |
\gamma|^2 \rangle$. Although this estimator has a high signal-to-noise
ratio and has been used extensively in the literature, it does not
separate the E- from the B-mode, unlike the aperture mass statistics.

Schneider et al.~(2002) calculated the covariance matrices $\Mat C$ of
the estimators defined in this section, in the case of a Gaussian
shear field. The covariances of the 2PCF consist of three terms: A
pure shot noise term (${\Mat D}$), originating from the dispersion of
the intrinsic ellipticities of the source galaxies, present only on
the diagonal, a cosmic variance term (${\Mat V}$), and a mixed term
(${\Mat M}$):
\begin{eqnarray}
\Cov_{++} \equiv \Cov(\hat \xi_+, \vt_1 ; \hat \xi_+, \vt_2) & = &
    {\Mat D} \, + \, {\Mat M}_{++} \, + \, {\Mat V}_{++} \nonumber \\
\Cov_{--} \equiv  \Cov(\hat \xi_-, \vt_1 ; \hat \xi_-, \vt_2) & = &
    {\Mat D} \, + \, {\Mat M}_{--} \, + \, {\Mat V}_{--}
    \label{Cov} \\
\Cov_{+-} \equiv \Cov(\hat \xi_+, \vt_1 ; \hat \xi_-, \vt_2) & = &
\phantom{{\Mat D} \, + \,} \, \, \, {\Mat M}_{+-} \, + \, {\Mat
V}_{+-}. \nonumber
\end{eqnarray}
The individual terms in full length are given in Schneider et
al.~(2002).  The covariance of both of the aperture mass statistics,
$\Cov({\cal M}_\pm)$, can be calculated from the 2PCF covariance
matrices using eq.~(\ref{Map-est}).

Derivatives of the covariances with respect to cosmological parameter
are easily obtained by differentiating eqs.\
(\ref{Cov}). The shot-noise term, being independent
of cosmology, does not contribute, and the cosmic variance term, being
quadratic in the correlation function, has a larger influence than the
mixed term.

The covariances of these estimators were studied in detail by
Schneider et al.~(2002), where covariance matrices were constructed
using an ensemble average over galaxy positions. Kilbinger \&
Schneider (2004) calculated the covariances directly by summing over
a specific realisation of galaxy positions which we follow here. This
has the advantage of being able to handle the discreteness resulting
from the finite galaxy number density in a correct way. Moreover,
incorporations of realistic survey strategies is done in a natural
way.

We note here that the cosmic variance term $\Mat V$ only contains the
Gaussian or unconnected part of the four-point correlator of shear. On
scales below $\sim$ 10 arc minutes, the non-Gaussianity of the shear
field gets important (van Waerbeke et al.~2002; Scoccimarro,
Zaldarriaga \& Hui 1999). On scales below $\sim$ 1 arc minute the
shot noise term $D$ dominates over $V$ (Kilbinger \& Schneider 2004),
thus under the Gaussian assumption we expect to slightly underestimate
the covariances in this transition regime between 1 and 10 arc
minutes.

\subsection{Fisher information matrix}
\label{sec:fisher}

The Fisher matrix (Kendall \& Stuart 1969; TTH) is defined as

\begin{equation}
{F}_{\alpha\beta} = \average{\frac{\del^2[-\ln \cal L]}{\del p_\alpha \del
    p_\beta}}
 = \left(\frac{\del^2[-\ln \cal L]}{\del p_\alpha \del
    p_\beta}\right)_{\Vec p = \Vec p_0},
\label{fisher-def}
\end{equation}
where the likelihood ${\cal L}$ depends on a vector of model
parameters $\Vec p$ $= (p_1, p_2, ...)$ and $\Vec p_0$ denotes the
true parameter values. The Fisher matrix is the
expectation value of the Hessian matrix of $(- \ln {\cal L})$ at $\Vec
p = \Vec p_0$ and is roughly spoken a measure of how fast ${\cal L}$
falls off from its maximum. The smallest possible variance $\Delta
p_\alpha$ of any unbiased estimator of some parameter $p_\alpha$ is given by the
Cram\'er-Rao inequality
\begin{equation}
\Delta p_\alpha \ge \sqrt{(F^{-1})_{\alpha\alpha}};
\label{mvb}
\end{equation}
the right-hand side of this inequality is called the minimum variance
bound (MVB). If only one parameter is to be determined from the data
and all others are fixed, the MVB simplifies to 
$(F_{\alpha\alpha})^{-1/2}$.

In the case of a Gaussian probability distribution function for the
parameters, the Fisher matrix becomes (Bunn 1995; Vogley \& Szalay
1996; TTH)
\begin{equation}
F_{\alpha\beta} = \frac 1 2 {\rm tr}\big[{\Mat C}^{-1} {\Mat C}_{,\alpha} \, {\Mat
  C}^{-1} {\Mat C}_{,\beta} + {\Mat C}^{-1} {\Mat M}_{\alpha\beta}\big].
\label{fisher}
\end{equation}
$\Cov_{,\alpha} \equiv \del \Cov/\del p_\alpha$ denotes the derivative of the
covariance with respect to the $\alpha^{\rm th}$ parameter. ${\Mat M}_{\alpha\beta}
= \Vec{ \mu}_{,\alpha} \Vec{\mu}_{,\beta}^{\rm t} + \Vec{\mu}_{,\beta}
\Vec{\mu}_{,\alpha}^{\rm t}$ where $\Vec \mu$ is mean of the data vector
$\Vec x$, $\langle \Vec x \rangle = \Vec \mu$. In our case, the
$n$-dimensional data vector $\Vec x$ consist of one or a combination
of the second-order statistics of shear as defined in Sect.\
\ref{sec:soss} (see also Sect.\ \ref{sec:data} for more details).

The $m=7$ parameters $p_\alpha$ (cosmological parameters plus source galaxy
redshift distribution characteristics) which we consider here are given in
Sect.~\ref{sec:cosmology}.

\subsection{Karhunen-Lo\`{e}ve (KL) eigenmodes and eigenvalues}

A general linear data compression can be written as
\begin{equation}
\tilde \Vec {x} = {\Mat T} \Vec x,
\label{data-compr}
\end{equation}
where from the $n$-dimensional data vector ${\Vec x}$ a new $\tilde
n$-dimensional data set $\tilde \Vec x$ is constructed via the $\tilde
n \times n$ matrix ${\Mat T}$. This means that the expectation value
$\langle{\Vec x}\rangle$ of the original data vector ${\Vec x}$
transforms to $\langle \tilde \Vec x \rangle = \Mat T \langle \Vec x
\rangle$ and hence the covariance $\Mat C= \langle \Vec x \Vec
x^\tp\rangle - \langle \Vec x \rangle \langle \Vec x \rangle^\tp$
changes accordingly to $\tilde {\Mat C} = {\Mat T}{\Mat C} {\Mat
  T}^\tp$. In our case, $\Vec x$ consists of the shear correlation
functions $\xi_+$ and $\xi_-$ or the aperture mass dispersion $\langle
\Mapsq \rangle$, representing $n$ data points measured for various
angular scales, see Sect.~\ref{sec:data}.  An example for a data
compression is already given in eq.~(\ref{Map-est}): $\langle \Mapsq
\rangle$ is a linearly compressed version of $\xi_+$ and $\xi_-$, the
compression matrix $\Mat T$ depends in this case on the functions
$T_+$ and $T_-$.

The Fisher matrix ${\Mat F}$ corresponding to the original
data vector ${\Vec x}$ changes to $\tilde {\Mat F}$, given by
\begin{eqnarray}
  \tilde{F}_{\alpha\beta} & = & {1 \over 2}{\rm tr}\left [ ({\Mat T \Cov \Mat
      T}^\tp)^{-1} ({\Mat T \Cov}_{,\alpha} {\Mat T}^\tp) ({\Mat T \Cov
      \Mat T}^\tp)^{-1} ({\Mat T \Cov}_{,\beta} {\Mat T}^\tp)
    \right. \nonumber\\ && \left. + ({\Mat T \Cov \Mat T}^\tp)^{-1}
    ({\Mat T \Mat M}_{\alpha\beta} {\Mat T}^\tp) \right]
  \label{fisher-prime}
\end{eqnarray}
Clearly, when the dimensionality $n$ and $\tilde n$ of the original and
the transformed data vectors are the same, the
transformation is a similarity transformation and the Fisher matrix
remains unchanged, $\tilde {\Mat F} = {\Mat F}$.

The goal of the KL analysis is to find a compression matrix ${\Mat T}$
for $\tilde n < n$ without loosing too much of information and to
estimate cosmological parameters with as small error bars as
possible. Considering the minimum variance bound (\ref{mvb}) of some
cosmological parameter $p_\alpha$, we seek to maximise the Fisher matrix
element $\tilde {F}_{\alpha\alpha}$.

Following TTH, we first consider the two simple cases of a constant
mean $\Vec \mu$ and a constant covariance ${\Mat C}$, where constant
means with respect to the parameter vector $\Vec p$. Then, we optimise
for several parameters simultaneously with the aid of a singular value
decomposition (SVD).

For cosmic shear, the assumption of a constant covariance seems quite
natural, since in most cases of weak lensing observations, the
covariance is not calculated analytically but extracted directly from
the data (e.g.\ van Waerbeke et al.~2002) and thus it is constant and
not depending on cosmological parameters. However, the situation is
different for the mean, since with a model parameter independent mean,
it is not possible to even define a likelihood function. And although
within the Fisher matrix formalism, a constant mean can be considered
in a consistent way, this mean first has to be found using e.g.\ the
maximum likelihood estimator. But the distinction into a constant mean
and a constant covariance case is of rather technical nature, and the
general case of parameter dependend mean and covariance can be
considered by combining the two, see Sect.~\ref{sec:general}. Thus, a
strategy to obtain cosmological parameters and their minimum variance
bound from data would be first to define a likelihood function $\cal
L$ using a parameter dependent mean (and parameter dependent
covariance if desired) to find the maximum likelihood parameter $\Vec
p_0$. Second, the Fisher matrix is calculated by evaluating the second
derivatives of $\cal L$ at the point $\Vec p = \Vec p_0$. A KL
analysis and data compression can be undertaken by first considering
the two cases of constant mean and covariance independently, and later
combining both cases.

Note that in our case, the ``mean'' is a second-order statistics (the
shear correlation), and the covariance is of fourth order (although
reduced to only depend on second-order because of the assumption of a
Gaussian shear field). We therefore choose a different approach than
e.g. Heavens (2003) or Seljak (1998) who directly use the galaxy
ellipticities as data vector. In their case, the (zero) mean is of
first order, and the covariance contains the information about the
power spectrum.

\subsubsection{Constant Mean}
\label{sec:const-mean}

In the case where the mean $\Vec \mu$ is known and constant, the
second term in (\ref{fisher-prime}) vanishes, since $\Vec
\mu_{,\alpha} = 0$ and therefore $\Mat M_{\alpha\beta}$ vanishes,
too. We first consider a compression matrix which consists of only one
row vector, ${\Mat T} = \Vec b^\tp$. The task of maximising the Fisher
matrix diagonal element $F_{\alpha\alpha}$ in order to minimise the error bar on
the cosmological parameter $p_\alpha$ is then equivalent to maximising
the eigenvalue $\lambda$ of the generalised eigenproblem ${\Mat
C}_{,\alpha} {\Vec b} = \lambda {\Mat C}{\Vec b}$. By
Cholesky-decomposing the symmetric and positive definite covariance
matrix ${\Mat C} = {\Mat L \Mat L}^\tp$, it can be reduced to the
ordinary eigenproblem
\begin{equation}
({\Mat L}^\tp {\Mat C}_{,\alpha} {\Mat L}^{-t}) {\Mat L}^\tp {\Vec b} = \lambda {\Mat
      L}^\tp {\Vec b}.
\label{evproblem}
\end{equation}
Solving this equation for all $n$ orthogonal eigenvectors ${\Mat
L}^\tp {\Vec b}_k$ gives us $n$ real eigenvalues $\lambda_k$. A
compression matrix ${\Mat T}$ is then constructed containing as row
vectors the first $\tilde n$ eigenvectors which have been sorted by
the absolute value of their corresponding eigenvalues. The matrix
$\Mat T$ represents a set of eigenvectors, rank-ordered according to
their signal-to-noise ratio.

The KL eigenmodes defined in this way satisfy an orthogonality
relation. Eigenmodes with high eigenvalues or low rank numbers contain
more information about a specific parameter, whereas the ones with
small (absolute) eigenvalues and high rank numbers contain almost no
additional information. As will be shown in Sect.~\ref{sec:KL-num},
we typically achieve a compression by a factor of nearly two for the
independent analysis of various parameters with getting the same MVB
as for the uncompressed case.

An individual eigenmode $\Vec b_k$ contributes to the measurement
error for the parameter $p_\alpha$ as $\delta p_\alpha = 1 / |\lambda_k|$. Thus,
the signal-to-noise ratio for this mode therefore is ${p_\alpha (\delta
p_\alpha)^{-1}} = p_\alpha |\lambda_k|$.

The new Fisher matrix diagonal element $\tilde F_{\alpha\alpha}$ corresponding to a KL
compression with $\tilde n$ eigenmodes is simply (TTH)
\begin{equation}
\tilde F_{\alpha\alpha} = \frac 1 2 \sum_{k=1}^{\tilde n} \lambda_k^2.
\label{tilde-F-cmean}
\end{equation}
In our analysis (Sect.\ \ref{sec:KL-num}), we repeat the data
compression, starting from the second-order statistics $\xi_\pm$ and
$\langle M_{\rm ap}^2 \rangle$ given for $n$ angular separations
and plot for various cosmological parameters $p_\alpha$ the associated
error $\Delta p_\alpha = 1/(\tilde F_{\alpha\alpha})^{1/2}$ as a
function of $\tilde n$, where $\tilde n \le n$ is the dimension of the
compressed data vector $\tilde \Vec {x}$, see Fig.\ 
\ref{fig:fisher+svd}. Although being a decreasing function of $\tilde
n$, the error reaches a constant plateau for some $\tilde n_0 < n$,
thus the original error for the uncompressed case is recovered before
all KL modes are used for the parameter estimation. We find in most of
the cases that a plateau is reached for $\tilde n_0 \lsim n/2$, thus a
compression factor of nearly two is possible without any loss of
information. This compression is with respect to the number of angular
separations $n$ at which the second-order shear statistics are
measured, representing a binned version of the data vector comprising
of all observed galaxy pairs. Note that the compression factor depends
of course on the original number of bins --- we comment on the binning in
Sect.~\ref{sec:bin}.

Since in the case of a constant mean, the Fisher matrix contains
products of the inverse covariance and the (derivative of the)
covariance, it is independent of the survey area, and very little
sensitive of the survey geometry.

\subsubsection{Constant covariance}
\label{sec:const-cov}

In contrast to the constant mean case, there is only one eigenvector
which contains all of the available information when the covariance
is known and independent of the parameters.  In this case, the first
term in (\ref{fisher-prime}) vanishes, and the corresponding eigenproblem is
\begin{equation}
({\Mat L}^{-1} {\Mat M}_{\alpha\alpha} {\Mat L}^{-t}) {\Mat L}^\tp {\Vec b} = \lambda {\Mat
    L}^\tp {\Vec b}
\end{equation}
which has only one non-trivial solution ${\Mat L}^\tp \Vec{b}_0 = {\Mat
L}^{-1} \Vec \mu_{,\alpha}$ with eigenvalue $\lambda_0 = 2 | {\Mat L}^{-1} \Vec
\mu_{,\alpha}| = \rm{tr} \left[ {\Mat C}^{-1} {\Mat M}_{\alpha\alpha} \right]$.

Inserting the only eigenvector $\Vec b_0 = \Mat C^{-1} \Vec \mu_{,\alpha}$
for the compression matrix $\Mat T$ into (\ref{fisher-prime}), we find
that the modified Fisher matrix is the same as the original one,
$\tilde F_{\alpha\beta} = F_{\alpha\beta} = \Vec \mu_{,\alpha}^{\rm t} \Cov^{-1} \Vec
\mu_{,\beta}$. This is not surprising since all data is collected
into one mode, and no information is lost.

If geometrical effects of the survey are neglected, the covariance is
proportional to one over the survey area $A$. Thus, in the case of
constant covariance the Fisher matrix is proportional to $A$, in contrast
to the constant mean case, where $\Mat F$ is independent of $A$. For
reasonable large cosmic shear surveys, the second term in
(\ref{fisher}) is therefore dominant over the first one (Sect.~\ref{sec:noise}).

\subsubsection{General case}
\label{sec:general}

TTH describe how the general case (when neither mean nor covariance is
constant) can be treated efficiently, by simply adding the one
eigenmode from the constant covariance case to the $\tilde n$ modes
from the constant mean analysis. However, as mentioned in the previous
section, the constant covariance eigenmode is (for reasonably large
survey areas) dominant over all the other modes and contains the bulk
part of the information about cosmological parameters. Thus, we do not
consider to combine these two cases.

\subsubsection{Joint parameter estimation}
\label{sec:svd}

\begin{figure}
\protect\centerline{
\epsfysize = 3.5 truein
\epsfbox[20 145 591 750]
{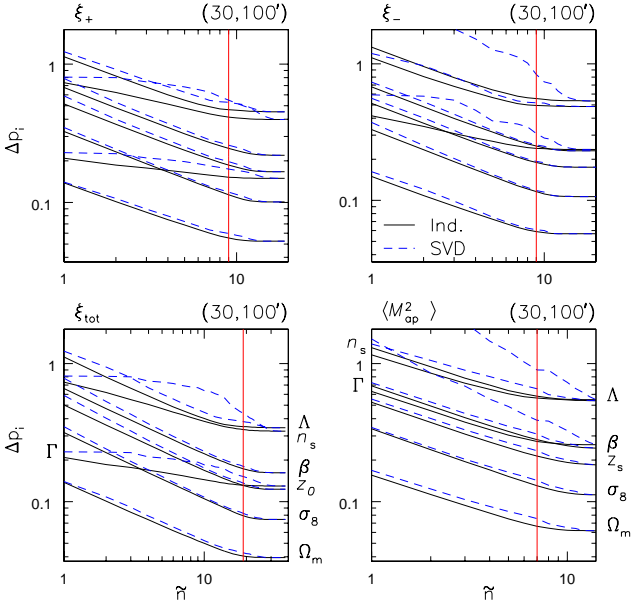}}
\caption{$\Delta p_\alpha$ = $(\tilde F_{\alpha\alpha})^{-1/2}$ as a
  function of the mode number $\tilde n$ is plotted, according to
  (\ref{tilde-F-cmean}) for the constant mean case and various
  cosmological parameters.  Solid lines correspond to the independent
  parameter estimation (Sect.~\ref{sec:const-mean}), dashed lines
  represent the joint parameter analysis (Sect.~\ref{sec:svd}). The
  four panels correspond to the shear statistics $\xi_+$, $\xi_-$,
  their combination $\xi_{\rm tot}$ and $\langle \Mapsq \rangle$ (see
  Sect.~\ref{sec:data}) The survey geometry is $(30,100^\prime)$.  The
  maximum mode number $n$ equals the number of bins, which is 38 for
  $\xi_{\rm tot}$, 19 for both $\xi_+$ and $\xi_-$ and 14 for $\langle
  \Mapsq \rangle$.
}
\label{fig:fisher+svd}
\end{figure}

For the independent estimation of a parameter $p_\alpha$, the KL
method is devised to minimise the associated error bar $\Delta
p_\alpha = (\tilde {F}_{\alpha\alpha})^{-1/2}$ in a suitably rotated
basis. However, for the case of joint parameter estimation, the object
to be minimised is the diagonal of $({\Mat F}^{-1})^{1/2}$ which is a
more demanding optimisation problem.  We follow TTH who discuss an
alternative approximate technique which virtually does the same. The
individual compression matrices ${\Mat T}_\alpha$, optimised for the
independent estimation of the parameter $p_\alpha$, are arranged into
a new matrix after multiplying each row (which corresponds to an
eigenvector) with its corresponding eigenvalue as ${\Mat T} =
(\Mat{\mathbf\Lambda}_1 {\Mat T}^\tp_1, \dots, \Mat{\mathbf\Lambda}_m
{\Mat T}^\tp_m)$.  Here, $\Mat{\mathbf\Lambda}_\alpha = {\rm
  diag}(\lambda_{\alpha 1}, \dots, \lambda_{\alpha n})$ is a diagonal
matrix containing the eigenvalues corresponding to the individual
eigenanalysis of the $\alpha^{\rm th}$ parameter.  ${\Mat T}$ will
contain a lot of redundant information, because of unavoidable
near-degeneracies between parameters, e.g.
between $\Omegam$ and $\sigma_8$ or between $\Gamma$ and $n_{\rm s}$.

In order to separate useful from redundant information, this new
matrix ${\Mat T}$ is factorised using a singular value decomposition
(SVD),

\medskip
\begin{raggedright}
\resizebox{\hsize}{!}{
  \includegraphics{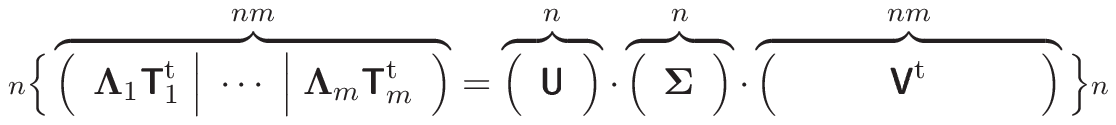},\hfill
}
\end{raggedright}

%
\noindent where ${\Mat U}^\tp
{\Mat U}$ = ${\Mat V}^\tp {\Mat V} = 1$ and ${\Mat {\mathbf\Sigma}} = {\rm
diag}(\sigma_i)$ is a diagonal matrix containing the singular values $\sigma_i$,
which can be interpreted as generalised eigenvalues. After sorting the
columns of ${\Mat U}^\tp$ according to the absolute values of the
corresponding $\sigma_i$
the final compression matrix is ${\Mat T}_{\rm joint} = {\Mat U}^\tp$.
Modes with high singular values contain the bulk information about the
cosmological parameters, whereas row vectors of ${\Mat T}_{\rm joint}$
corresponding to vanishing or very small $\sigma_i$ capture
redundant or almost redundant information. Inserting ${\Mat T}_{\rm
  joint}$ into (\ref{fisher-prime}), the Fisher matrix after the final
compression is easily calculated.

If {\EDIT $\tilde n_\alpha$} is the number of modes that were used in
the optimisation process for the $\alpha^{\rm th}$ parameter, the
joint matrix ${\Mat T}_{\rm joint}$ has $\tilde n^\prime =
\sum_{\alpha=1}^m \tilde n_\alpha$ columns. Based on the amplitude of
the singular values, one can choose a final mode number $\tilde n_{\rm
  joint} < \tilde n^{\prime}$ fixing the compression factor. As an
example, the transformed diagonal elements of the Fisher matrix are
plotted as a function of $\tilde n_{\rm joint}$ in Fig.\ 
\ref{fig:fisher+svd}, where the saturation plateau can be used to
determine the final mode number.


Also for the constant covariance case (Sect~\ref{sec:const-cov}), a
joint parameter estimation analysis can be done. We combine the $m$
eigenvectors resulting from the $m$ individual parameter analyses to
the $n \times m$-matrix $\Mat T$, after having multiplied
them by their respective eigenvalues. The singular value decomposition
of $\Mat T$ will result in $m$ singular values, which give us
information of the degree of degeneracy between the parameters. We
will discuss results for this analysis in Sect.~\ref{sec:KL-num}.

\subsubsection{Window functions}

The linear data compression defined in (\ref{data-compr}) generates a
new data vector $\tilde \Vec x$ which in the case of single parameter
optimisation represents a set of uncorrelated unit-variance Gaussian
random variables (TTH). The components $\tilde x_i$ of the new data vector
contain pairwise independent and uncorrelated information about
cosmology. In our case the data consists of second-order shear
statistics and the cosmological information is contained in the
convergence power spectrum.

Using the KL eigenmode technique, we can study in detail how the power
spectrum is sampled in order to yield uncorrelated data points. By
comparing modes corresponding to high (low) eigenvalues, we can see
which scales carry much (little) of the cosmological information.

For any second-order statistics, the dependence on the power spectrum
is encoded in a window function. For the two shear 2PCF $\xi_+$ and
$\xi_-$, these are the broad-band Bessel functions $\J_0$ and $\J_4$,
respectively, see eq.~(\ref{xi-pm-def}); the filter function for the
aperture mass statistics $\langle\Mapsq\rangle$ is the narrow-peaked
function $[24 \, \J_4(\eta)/\eta^2]^2$, see (\ref{Map-disp}).

For each component $\tilde x_i$ of the new data vector, we define a new
window function $W_i$, which is a combination of the original filter
functions associated to different smoothing scales $\theta_j$. In
the case of the 2PCF, the new data vector is
\begin{eqnarray}
\tilde \xi^\pm_i = T^\pm_{ij} \xi_\pm(\theta_j) = {1 \over 2 \pi}
 \int_0^\infty \dd \ell \, \ell \, P_{\kappa}(\ell) W_i^{\pm}(\ell); \nonumber \\ 
 W_i^{\pm}(\ell) = \sum_j  T^{\pm}_{ij} \J_{0,4} (\ell \theta_j).
\label{integral}
\end{eqnarray}
Similarly, for $\langle\tilde\Mapsq\rangle$ we obtain the new window
functions $W_i^{\rm E}(\ell) = \sum_j T^{\rm E}_{ij} [24 \, \J_4(\ell
\theta_j)/(\ell \theta_j)^2]^2$. In the case of the combined vector
$\tilde \xi^{\rm tot} = (\tilde \xi^+, \tilde \xi^-)$, the window
functions $W_i^{\rm tot}$ are linear combinations of $W_i^+$ and
$W_i^-$.

In case of a constant mean, the rows of the matrix $\Mat T$ contain
the $\tilde n$ transposed eigenvectors $\Vec b$, see
Sect.~\ref{sec:const-mean}. For the constant covariance case, there is
only one eigenmode, in this case $\Mat T_0 = \Vec \mu_{,\alpha}^{\rm
  t} \Mat C^{-\rm t}$. We denote all quantities corresponding to the
constant covariance case with a subscript `0', e.g.\ $\tilde \xi_0^+$,
$W^+_0$ etc. For the joint parameter estimation
(Sect.~\ref{sec:svd}), window functions are defined analogously, by
inserting the joint matrix $\Mat T_{\rm joint}$ into
eq.~(\ref{integral}).

The study of these window functions, which is presented in the next
section, will provide us with insights about how the
convergence power spectrum is sampled in order to constrain
cosmological parameters with cosmic shear. It allows us to investigate
how different Fourier modes are probed by a given survey strategy.

\section{Numerical Results}
\label{sec:numerical}

The impact of various survey strategies, noise structures and the
effect of binning are considered in the determination of the
eigenmodes. We study separately the constant
mean case (see Sect.\ \ref{sec:const-mean}) and the constant covariance
case (see Sect.\ \ref{sec:const-cov}). Eigenmodes associated with the
independent analysis of a single parameter and the joint analysis of
several parameters (see Sect.~\ref{sec:svd}) are investigated.

\subsection{The input data}
\label{sec:data}

Our input data vector $\Vec x$ is one of one the following
second-order statistics of cosmic shear, as defined in Sect.\
\ref{sec:soss}:

\begin{enumerate}

\item[1.]  The correlation function $\xi_{+}(\theta_i)$ measured at
various angular scales $\theta_i$.  The covariance is denoted as
$\Cov_{++}$.

\item [2.]  The correlation function $\xi_{-}(\theta_i)$ measured at
various angular scales $\theta_i$.  The covariance is denoted as
$\Cov_{--}$.

\item [3.] The joint combination $\xi_{\rm tot}$ of both data vectors
$\xi_{+}$ and $\xi_{-}$ where the joint
covariance matrix $\Cov_{\rm tot}$ is constructed out of blocks
$\Cov_{++}$, $\Cov_{--}$ on the diagonal and $\Cov_{+-}$ and its
transpose on the off-diagonal.

\item [4.] The aperture mass statistics
  $\langle\Mapsq(\theta_i)\rangle$ for various aperture radii
  $\theta_i$ and its covariance $\Cov({\cal M}_{+})$.

\end{enumerate}

These statistics are calculated from the convergence power spectrum
using eqs.~(\ref{xi-pm-def}) and (\ref{Map-disp}). For the 2PCF, we
use 19 angular logarithmic bins, the smallest separation between two
galaxies considered being 0.5 arcmin. Since the calculation of the
covariances and their derivations are very time-consuming, we have
chosen a rather small number of angular bins. The largest angular
distance is determined by the survey geometry (Sect.\
\ref{sec:survey_char}) -- for a patch geometry, we choose this to be
slightly less than the patch diameter $R$. This is to avoid very large
separations with small numbers of galaxy pairs, which caused our
results to be instable when we varied the binning.

The joint 2PCF data vector $\xi_{\rm tot}$ has 38 entries,
corresponding to the two times 19 angular bins for $\xi_+$ and $\xi_-$. We
calculate the aperture mass $\langle\Mapsq\rangle$ for 14 different
angular bins. The calculation of the covariance of
$\langle\Mapsq\rangle$ as a weighted sum of the 2PCF covariances can
be biased for small aperture radii, when for a given binning of the
2PCF only a small number of terms contribute to the sum. We take into
account a smaller number of aperture radii than separations for the
2PCF.

\subsection{Cosmological model and parameters}
\label{sec:cosmology}

We include $m=7$ parameters in our analysis. These are $\Omega_{\rm
m}$, the power spectrum normalisation $\sigma_8$, the spectral index
of the initial scalar fluctuations $n_{\rm s}$ and $\Gamma$ which
determines the shape of the power spectrum. In addition, the
cosmological constant $\Lambda$ is included in our analysis. The
redshift distribution of source galaxies is assumed to be (Smail et al.~1995)
\begin{equation}
p(z) \dd z = { \beta \over z_0 \Gamma(3/\beta) } \left ({ z \over z_0}
\right)^2 e^{-(z/z_0)^{\beta}} \dd z.
\label{pzdz}
\end{equation}
The parameters $z_0$ and $\beta$ are treated as free parameters
to be determined from the data.

Our reference model is a flat $\Lambda$CDM Universe with $\Omegam =
0.3$, $\Gamma = 0.21$, $n_{\rm s} = 1$ and $\sigma_8 = 1$. The source
redshift distribution is given by $z_0 = 1$ and $\beta = 1.5$,
corresponding to a mean redshift of $\sim 1.5$. The CDM transfer
function is taken from Bardeen et al.~(1986), the non-linear power
spectrum is calculated using the fitting formulae of Peacock \& Dodds
(1996).

\subsection{KL eigenmode analysis}
\label{sec:KL-num}

\subsubsection{Error bars}
\label{sec:errorbars}

For the case of a constant mean (Sect.~\ref{sec:const-mean}), we plot
the MVB $\Delta p_\alpha = (\tilde F_{\alpha\alpha})^{-1/2} =
\big(2/\sum_{k=1}^{\tilde n} \lambda_k^2\big)^{1/2}$ in
Fig.~\ref{fig:fisher+svd} as a function of the compression mode number
$\tilde n$, for both single and joint parameter estimation. Typically,
the asymptotic limit is reached at a lower mode number in case of
$\tilde \xi^{+}$ compared to $\tilde \xi^{-}$.  On the other hand, the
joint use of $\tilde \xi^{+}$ and $\tilde \xi^{-}$ reduces the error
bars by a factor of nearly two in most cases. The saturation limit is
approximately the same for each of the individual cosmological
parameters. Irrespective of the survey strategy, approximately the
first half of eigenmodes contain virtually all information about each
individual parameter, thus a compression factor of nearly two is
possible without increasing the error.

The error bars attained by both the individual and the joint analysis
of the 2PCF $\xi_{+}$ and $\xi_{-}$ are typically tighter compared to the
$\langle \Mapsq \rangle$-constraints for a given survey geometry.

We also conduct a joint parameter estimation for the case of a
constant covariance (Sect.~\ref{sec:const-cov}). The complete
information of all seven cosmological parameters is encoded in the
$m=7$ individual eigenmodes. However, the result of the SVD shows,
that for $\Omegam, \sigma_8, z_0$ and $\beta$ already the first
singular mode carries basically all of the information on these
parameters. For $\Gamma, n_{\rm s}$ and $\Lambda$, the saturation of
the error is reached after two modes. Apparently, the first mode picks
up most of the information about the first group of highly-degenerate
parameters, the second one completes the information about the
parameters from the second group. This picture is consistent with the
correlation matrix of the parameters, as discussed in
Sect.~\ref{sec:corr-coeff}.

\subsubsection{Window functions for the compressed eigenmodes}

\begin{figure}
\protect\centerline{
\epsfysize = 3.6truein
\epsfbox[20 146 591 715]
{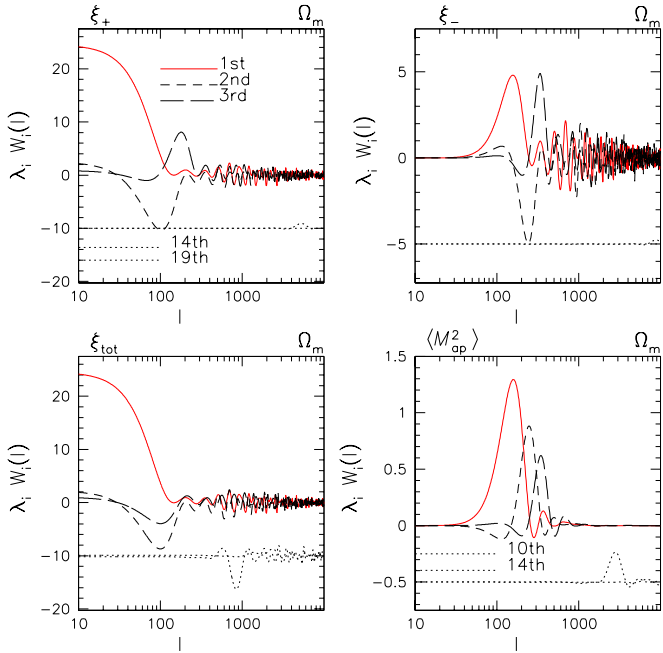}}
\caption{Window functions $W_i(\ell)$ multiplied with the
  corresponding eigenvalues $\lambda_i$, associated with the
  individual determination of the parameter $\Omegam$, for the
  constant mean case.  The first three eigenmodes, representing the
  three highest signal-to-noise modes, and for comparison, two exemplary
  higher order eigenmodes ($i=14,19$ for $\xi_+, \xi_-$ and $\xi_{\rm
    tot}; \; i=10,14$ for $\langle \Mapsq \rangle$) are plotted. The
  four panels correspond to the four different shear statistics as
  indicated. The specific survey strategy used in this case is a $(30,
  100^\prime$) patch geometry (see Sect.~\ref{sec:survey_char}). Note that
  the window functions for the two higher order modes have been
  displaced from zero to increase the visibility.}
\label{fig:eigen_omega}
\end{figure}

In Fig.~\ref{fig:eigen_omega}
we plot the rank-ordered window functions $W_i(\ell)$ associated with
$\Omegam$
for the constant mean case. As an example, the first three window
functions containing most of the information are compared with two
higher order modes which contain less or negligible information.
The survey strategy
used in this analysis is a $(30, 100^\prime$)-patch geometry (see
Sect.~\ref{sec:survey_char}).

\begin{figure}
\protect\centerline{
\epsfysize = 3.6truein
\epsfbox[20 146 591 710]
{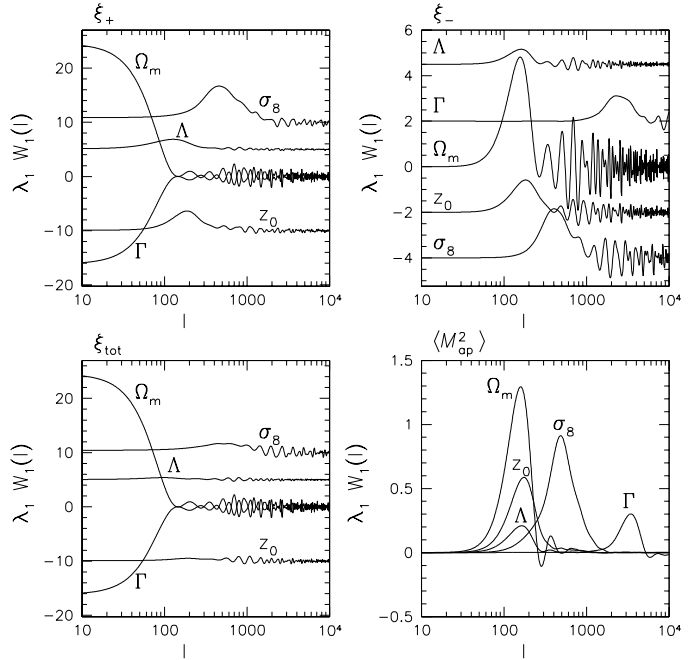}}
\caption{The dominant first window functions $\lambda_1 W_1(\ell)$ for
the constant mean case, associated with various parameters as indicated
in the figure. The four panels correspond to the four different shear
statistics as indicated. The survey strategy is $(30, 100^\prime$)
(see Sect.~\ref{sec:survey_char}). For visual clarity, some of the curves are
shifted vertically.}
\label{fig:eigen_mean}
\end{figure}

\begin{figure}
\protect\centerline{
\epsfysize = 3.6truein
\epsfbox[20 146 591 715]
{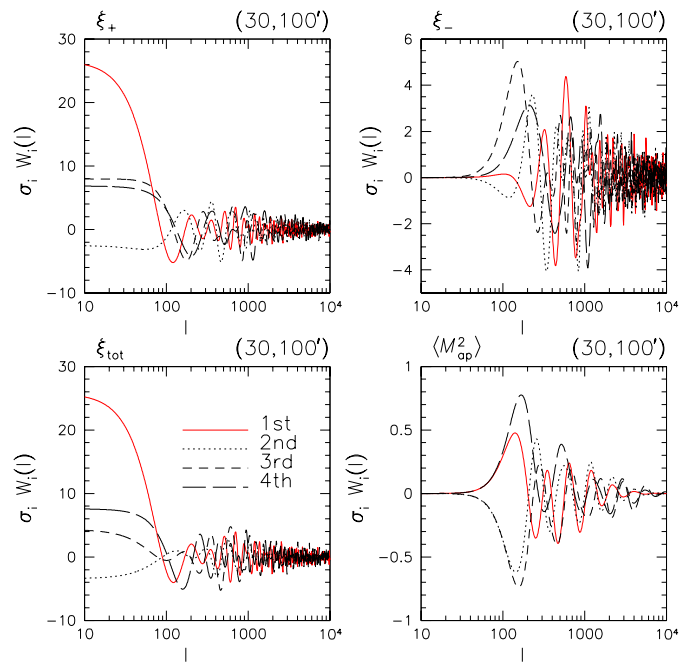}}
\caption{Window functions $\sigma_i W_i$ for the joint analysis of all
  parameters using SVD in the case of constant mean. The first four
  singular modes multiplied by the corresponding singular values are
  displayed.}
\label{fig:Wsvd}
\end{figure}

\begin{figure}
\protect\centerline{
\epsfysize = 3.6truein
\epsfbox[20 146 591 715]
{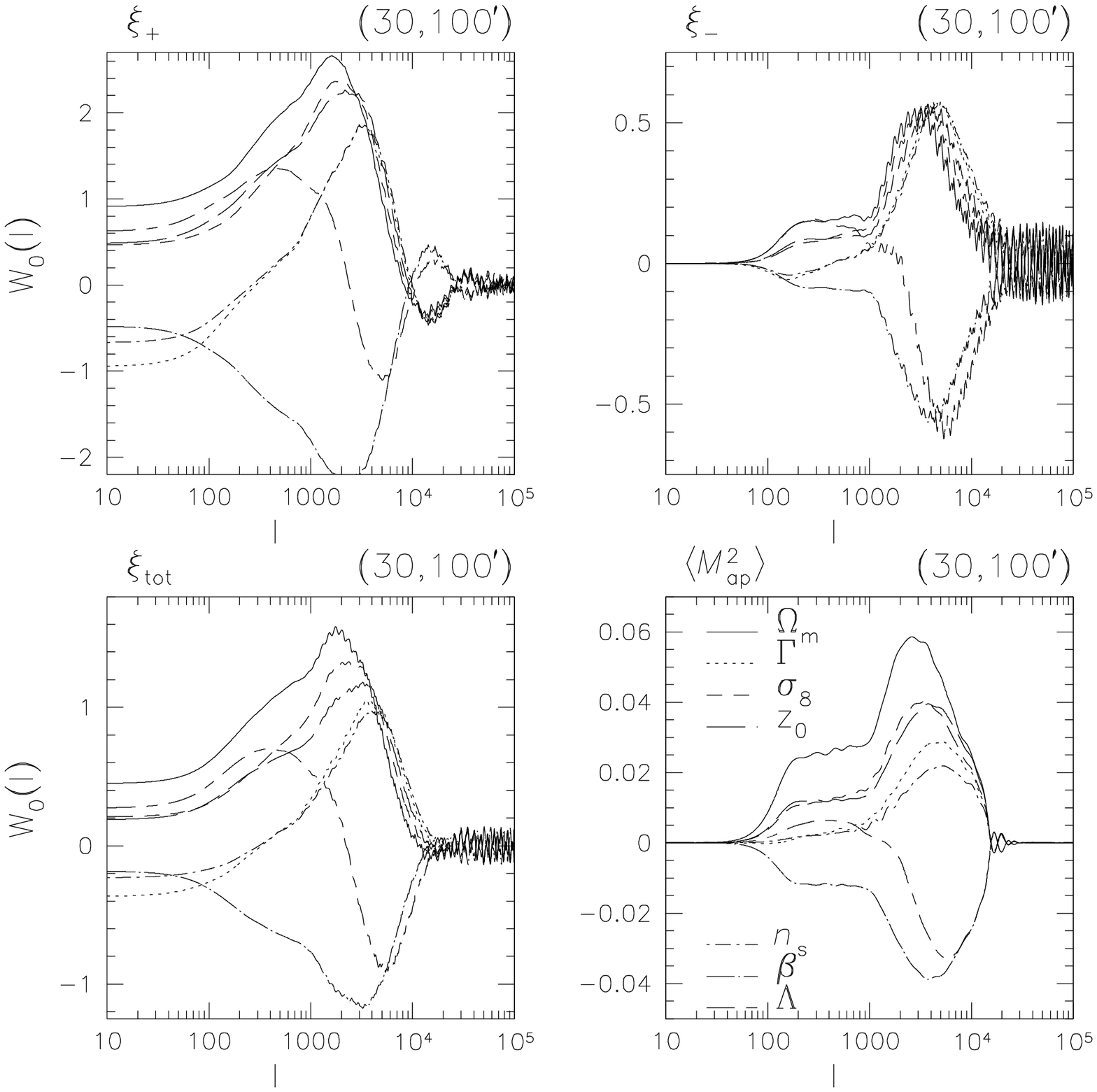}}
\caption{The window function $W_0$ for the constant covariance case
associated with various parameters as indicated in the figure. The
four panels correspond to the four different shear statistics as
indicated.}
\label{fig:W0}
\end{figure}

Because the window functions for $\tilde \xi^+_{i}$ are linear
combinations of $\J_0(\ell\theta_j)$, there is always an extended tail
which takes contributions from very large scales. The window functions
for $\langle \tilde \Mapsq \rangle$ are very locallized and each
eigenmode samples only a small $\ell$-region. For all statistics,
small scales are noise dominated and contribute with only small
amplitude to higher-order, low signal-to-noise KL modes, see also Fig.~\ref{fig:peak}.


The fact that the information content is larger for small $\ell$ and
decreases with $\ell$ can be understood when considering the simple
case in which the covariance matrices and its derivatives are
diagonal. Then, the eigenproblem matrix $\Mat A = \Mat L^{\rm t} \Mat
C_\alpha \Mat L^{-\rm t}$ (\ref {evproblem}) is also diagonal with
$A_{kk} = (C_{,\alpha})_{kk} / C_{kk}$ being the eigenvalues. The
diagonals of both $\Mat C$ and $\Mat C_{\alpha}$ are decreasing
functions of the angular scale (except for small bumps due to
geometrical survey effects). Since the decrease of $\Mat C_{\alpha}$
is in general shallower than $\Mat C$, the largest eigenvalues occur
on the largest angular scales.

For the constant covariance case (Fig.~\ref{fig:W0}), the window
functions $W_0$ are much broader since there is only one eigenmode
containing all information and taking contributions from all angular
scales. The $W_0$ peak at a median angular scale where the signal
dominates over the noise both from the intrinsic ellipticity
dispersion of galaxies at small angular scales and from the finite sky
coverage at larger angular scales.

The strong degeneracy among various parameters is reflected in the
behaviour of their associated window functions: for all four
statistics, degenerate parameters have very similar curves (see
next section for a discussion of the different near-degeneracies). The
$\Gamma$- and $n_{\rm s}$-filter functions have a zero-transition --
the low $\ell$-plateau has opposite sign than the peak reflecting the
sensitivity of these parameters to a tilt in the power spectrum.

\begin{figure}
\protect\centerline{
\epsfysize = 3.5truein
\epsfbox[23 146 591 716]
{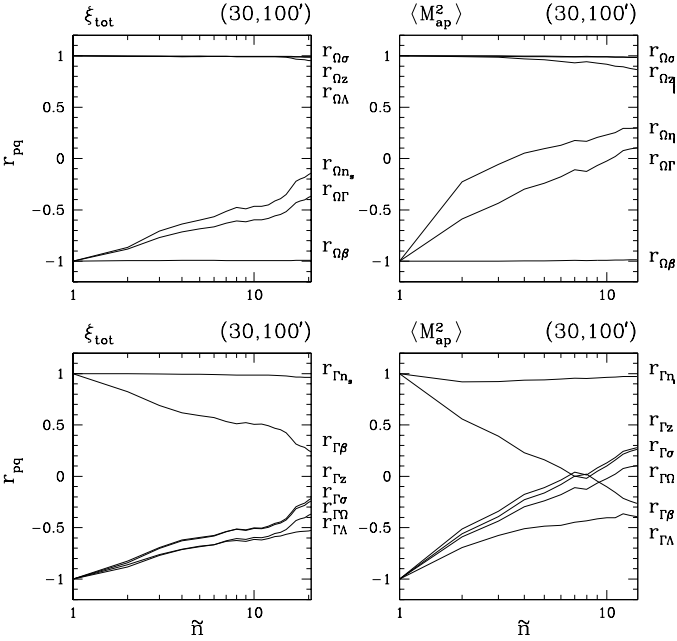}}
\caption{The correlation of $\Omegam$ (upper panels) and $\Gamma$
    (lower panels) with the other parameters, as a function of the
    mode number $\tilde n^\prime$ for the joint parameter estimation
    involving all parameters. The left and the right panels correspond
    to $\xi_{\rm tot}$ and $\langle\Mapsq\rangle$, respectively.}
\label{fig:r-omega}
\end{figure}

Figure \ref{fig:W0} displays for the dominant contibution to the
Fisher matrix (second term in eq.~\ref{fisher}) how individual Fourier
modes are sampled and combined in an ``optimal'' way to constrain
cosmological parameters as indicated. It shows on which
scales the convergence power spectrum has the largest influence on the
determination of cosmological parameters, depending on the survey
strategy (see also Sect.~\ref{sec:survey_strat}). In a forthcoming paper,
analogous analyses based on the present work will be performed for
third-order statistics of cosmic shear, allowing one to quantify the
sampling of the convergence bispectrum to extract cosmological
information. Moreover, the scales of interest will be studied for a
combined analysis of the power and the bispectrum which will reduce
the near-degeneracies between parameters.

\subsubsection{Correlation between cosmological parameters}
\label{sec:corr-coeff}

In Fig.~\ref{fig:r-omega} we plot the correlation coefficient of the
Fisher matrix $r_{\alpha\beta} = \tilde F_{\alpha\beta}/({\tilde
  F_{\alpha\alpha} \tilde F_{\beta\beta}})^{1/2}$, which quantifies
the level of degeneracy between two parameters, as
a function of the compression mode number $\tilde n$. For very small
mode numbers, the Fisher matrix becomes singular and all parameters
are completely degenerate. Clearly, there exist two groups of
highly-degenerate parameters ($\Omegam, \sigma_8, z_0$, $\beta$,
$\Omega_\Lambda$) and ($\Gamma$, $n_{\rm s}$). For parameters from different
groups, the correlation decreases, even for mode numbers $\tilde n$
where the diagonal element has reached the plateau. Thus, redundant
modes which do not carry any information regarding the Fisher diagonal
are nevertheless important and help to reduce parameter degeneracies.




We also calculate the correlation coefficient $r_{\alpha\beta}$ for
the constant covariance case, where the SVD gives us only $m=7$
singular modes. The correlation coefficient forms a plateau when as
few as two of the singular modes are included, as it is the case for
the diagonal elements of the Fisher matrix
(Sect.~\ref{sec:errorbars}). However, even though $\Mat F$ does not
change much when three modes and more are added, this is not true for
the inverse Fisher matrix since $\Mat F$ is quite ill-conditioned and
small changes have a large impact on $\Mat F^{-1}$. For a numerically
stable determination the MVBs from the inverse Fisher matrix, one
needs as many modes as number of parameters.


\subsubsection{Survey strategy}
\label{sec:survey_strat}

\begin{figure}
\protect\centerline{
\epsfysize = 1.7truein
\epsfbox[23 434 591 716]
{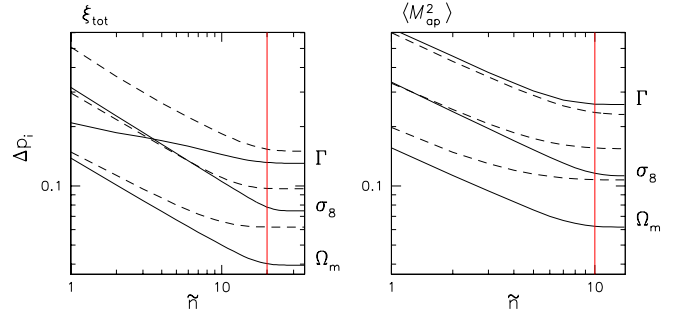}}
 \caption{$\Delta p_\alpha$ as function of the eigenmode number $\tilde n$,
  for the survey geometries $(30,100^\prime)$ (solid lines) and
  $75\cdot 26^{\prime \, 2}$ (dashed lines). The single parameter case
  of $\Omegam, \sigma_8$ and $\Gamma$ is displayed.}
\label{fig:beta-survey}
\end{figure}

We plot $\Delta p_\alpha$ as a function of the eigenmode number in
Fig.~\ref{fig:beta-survey}. In contrast fo Fig.~\ref{fig:fisher+svd},
where we compared the single and the joint (SVD) parameter case for
one survey strategy, we emphasis here on the single parameter
estimation and compare the resulting errors for different survey
geometries.  All patch geometry strategies as well as the survey
consisting of 12 uncorrelated $65^{\prime \, 2}$-fields yield very
similar results, the case $(30,100^\prime)$ is shown as a
representative. On the contrary, the MVBs from the $75\cdot 26^{\prime
  \, 2}$-geometry are significantly higher with few exceptions. This
latter survey does not sample medium and large scales in contrast to
the patches and the $12 \cdot 65^{\prime \, 2}$ case. The small cosmic
variance corresponding to the 75 independent lines of sight can not
compensate for the lack of large-scale information.

\begin{figure}
\protect\centerline{
\epsfysize = 3.5truein
\epsfbox[23 146 591 716]
{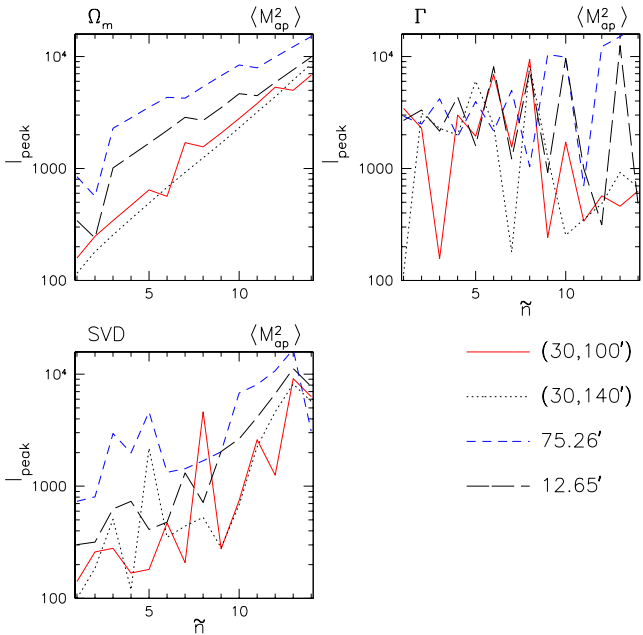}}
 \caption{The peak position of the $\tilde n^{\rm th}$ $\langle \tilde
\Mapsq \rangle$-windows as a function of the corresponding eigenmode
number $\tilde n$, for single analysis of the parameters $\Omegam$
(upper left panel) and $\Gamma$ (upper right), and for the joint
estimation (lower panel). $\theta_{\rm peak} = 5/\ell_{\rm peak} =
17^\prime (1000/\ell_{\rm peak})$ is the corresponding peak in real space.}
\label{fig:peak}
\end{figure}

\begin{figure}
\protect\centerline{
\epsfysize = 2.0truein
\epsfbox[23 407 591 716]
{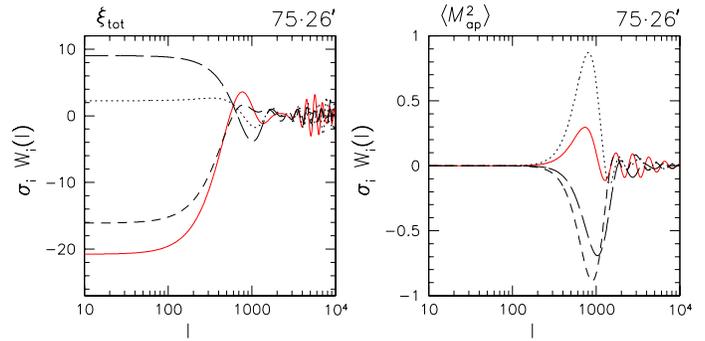}}
 \caption{The first four window functions $\sigma_i W_i$ for the joint
   parameter and constant mean case, corresponding to a $75\cdot
   26^\prime$ survey. See Fig.~\ref{fig:Wsvd} for an explanation of
   the curves, and for comparison with a $(30,100^\prime)$ geometry.}
\label{fig:Wsvd-survey}
\end{figure}

\begin{figure}
\protect\centerline{
\epsfysize = 2.0truein
\epsfbox[23 407 591 716]
{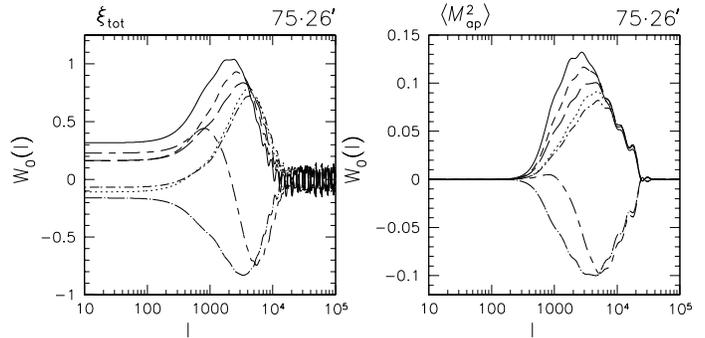}}
 \caption{The constant covariance window functions $W_0$ for the seven
   cosmological parameters, corresponding to a $75\cdot
   26^\prime$ survey. See Fig.~\ref{fig:W0} for an explanation of
   the different curves, and for comparison with a $(30,100^\prime)$ geometry.}
\label{fig:W0-survey}
\end{figure}

The effect of survey strategy on the window functions is a shift in
$\ell$, depending on the scales that are sampled by the survey
(compare Fig.~\ref{fig:Wsvd} with \ref{fig:Wsvd-survey}, and
Fig.~\ref{fig:W0} with \ref{fig:W0-survey}). For the constant mean
case, the uncorrelated images-surveys shift the peaks of the $\langle
\tilde \Mapsq \rangle$-window functions corresponding to $\Omegam$ and
the other parameters from this degeneracy group
(Sect.~\ref{sec:corr-coeff}) towards higher $\ell$ whereas for
$\Gamma$ and $n_{\rm s}$, the peaks seem to be randomly distributed,
see Fig.~\ref{fig:peak}. For the latter two parameters, there is no
preference of large scales, apparently all scales contribute with the
same importance to the Fisher matrix.

\begin{figure}
\protect\centerline{
\epsfysize = 3.5truein
\epsfbox[25 144 591 715]
{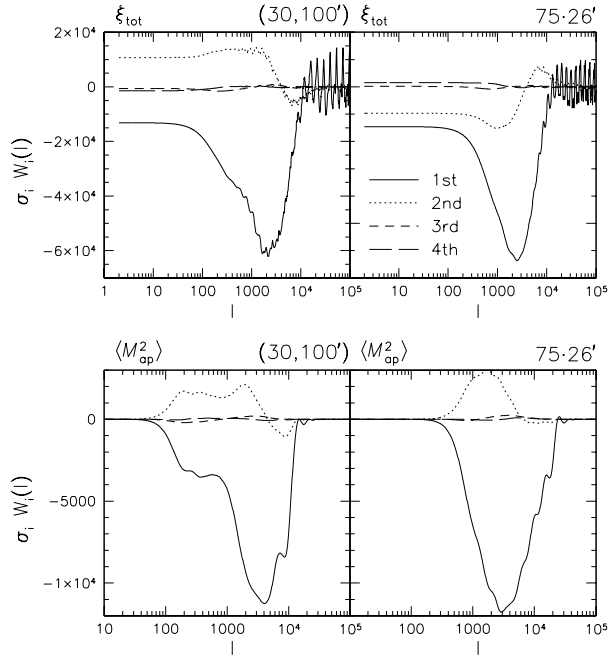}
}
\caption{The first four window functions multiplied by the singular
  values, $\sigma_i W_i, i=1\ldots 4$, for the joint analysis of constant
  covariance, corresponding to two survey strategies as indicated.}
\label{fig:Wsvd0}
\end{figure}

Figure \ref{fig:Wsvd0} shows the window functions $W_0$ for the joint
parameter analysis, corresponding to the dominant constant covariance
case. From this figure, it is clear that only the first two modes
carry significant information about cosmology. The shapes of the two
functions is complementary, for example, the second one has a zero
transition at roughly the peak position of the first one.  The reason is
that there are basically two groups of parameters entering the
convergence power spectrum, and the information content is
described by two independent singular modes.

For various survey strategies, we calculate the simultaneous MVBs for
$\Omegam, \sigma_8, \Gamma$ and $n_{\rm s}$ using (\ref{mvb}). The
patch geometries yield similar errors which are smaller than those
from the uncorrelated images-surveys, see Table \ref{tab:mvb-full}. In
particular for $\langle M_{\rm ap}^2 \rangle$, the lack of large
scales of the $75\cdot26^{\prime \, 2}$ case results in a large
uncertainty for $\Omegam$ and $\sigma_8$. This confirms a result found
by Jain \& Seljak (1997), stating that in order to break the
$\Omegam$-$\sigma_8$-degeneracy, the shear correlation on large
(linear) scales has to be added to the measurement of small (non-linear) scale
correlation.

\begin{table}

\label{tab:mvb-full}
\caption{Simultaneous MVBs for various parameters and survey strategies.}

\begin{center}
\begin{tabular}{ccccc}
\multicolumn{5}{c}{$\xi_{\rm tot}$} \\ \hline\hline
         & $\Omegam$ & $\Gamma$ & $\sigma_8$ & $n_{\rm s}$ \\ \hline 
$(60,140^\prime)$  &     0.13 & 0.17 & 0.24 & 0.21 \\
$(30,140^\prime)$  &     0.14 & 0.17 & 0.26 & 0.21 \\
$(60,100^\prime)$  &     0.14 & 0.19 & 0.26 & 0.24 \\
$(30,100^\prime)$  &     0.14 & 0.18 & 0.26 & 0.22 \\
$12\cdot65^{\prime \, 2}$ &     0.17 & 0.22 & 0.30 & 0.27 \\
$75\cdot26^{\prime \, 2} $ &     0.18 & 0.23 & 0.33 & 0.26 \\ \hline
& & & & \\
 \multicolumn{5}{c}{$\langle M_{\rm ap}^2 \rangle$} \\ \hline\hline
         & $\Omegam$ & $\Gamma$ & $\sigma_8$ & $n_{\rm s}$ \\ \hline 
$(60,140^\prime)$  &     0.22 & 0.35 & 0.41 & 0.45\\
$(30,140^\prime)$  &     0.27 & 0.40 & 0.49 & 0.50\\
$(60,100^\prime)$  &     0.25 & 0.43 & 0.47 & 0.54\\
$(30,100^\prime)$  &     0.28 & 0.46 & 0.53 & 0.57\\
$12\cdot65^{\prime \, 2}$ &     0.28 & 0.47 & 0.53 & 0.56\\
$75\cdot26^{\prime \, 2}$ &     0.37 & 0.53 & 0.69 & 0.58\\ \hline
\end{tabular}
\end{center}

\end{table}

\subsubsection{Noise levels}
\label{sec:noise}

We determine the dependence of the Fisher matrix for different survey
characteristics and noise levels. The calculations are done for a
($30,100^\prime$)-patch survey geometry. We vary the survey area $A$
between 1.4 and 14 square degrees, by adding more and more
independent, uncorrelated patches, each containing 30 images. For the
variation of the other parameters (source ellipticity dispersion
$\sigma_\ve$, number of background galaxies $n_{\rm gal}$, number of
bins $n$ and redshift parameter $z_0$), a single patch was
used to calculate the Fisher matrix. In the constant covariance case,
we write the MVB as
\begin{eqnarray}
\Delta p_\alpha =
F_{\alpha\alpha}^{-1/2} & \!\! = \!\! & k \, \left(\frac{A}{10 \, {\rm
sq \, deg}^2}\right)^{-0.5}
\left(\frac{\sigma_\ve}{0.3}\right)^{\mu}   \nonumber \\
& & \times \left(\frac{n_{\rm gal}}{30 \, {\rm arcmin}^{-2}
}\right)^{\nu}
\left( \frac{n}{20} \right)^\eta
z_0^{\lambda},
\label{MVB-fit}
\end{eqnarray}
where the constant $k$ and the power-law indices $\mu, \nu$, $\eta$ and
$\lambda$ are given in Table \ref{tab:mvb} for each parameter $\alpha$.

Note that although the MVB for $\langle \Mapsq \rangle$ is smaller
than the one for the 2PCF, this is not true when more than one
parameter is considered. In that case, the MVB is given by the
inverse of (a submatrix of) the Fisher matrix. The off-diagonal terms
are typically larger for  $\langle \Mapsq \rangle$ which decreases the
diagonal of the inverse matrix.

If the mean is constant, the Fisher matrix is independent of the area
of the survey. Except for the binning (see next section), the
sensitivity of the other survey characteristics is weaker than for the
constant covariance case. For our fiducial values of $\sigma_\ve =
0.3, n_{\rm gal} = 30 \, {\rm arcmin}^{-2}, n = 20$ and $z_0
= 1$, the survey area where both terms in (\ref{fisher}) are equal varies
between 0.06 (for $n_{\rm s}$) and 0.4 ($\Omegam$) square degrees in
the case of $\xi_{\rm tot}$. For the aperture mass statistics, this
area of equal contribution of both terms to the Fisher matrix is
roughly a factor of 4 smaller. Thus, only for very small survey areas
the constant mean term is important.

\begin{table}
  \caption{The coefficients describing the MVB for the constant
    covariance case as function of various survey characteristics, for
    different cosmological parameters, see (\ref{MVB-fit}).}
\label{tab:mvb}
\begin{tabular}{lrrrrr}
\multicolumn{6}{c}{$\xi_{\rm tot}$} \\ \hline\hline
Parameter & $k$ & $\mu$ & $\nu$ & $\eta$ & $\lambda$ \\ \hline
 $\Omegam$  & 0.0074  & $-$0.77 &  0.40 & $-$0.03 &  0.52 \\
  $\Gamma$  & 0.0109  & $-$1.28 &  0.71 & $-$0.08 &  1.03 \\
$\sigma_8$  & 0.0098  & $-$0.96 &  0.52 & $-$0.05 &  0.85 \\
     $z_0$  & 0.0157  & $-$1.08 &  0.59 & $-$0.06 & $-$0.59 \\
$n_{\rm{s}}$  & 0.0237  & $-$1.34 &  0.76 & $-$0.09 &  1.04 \\
   $\beta$  & 0.0206  & $-$1.09 &  0.59 & $-$0.06 &  0.36 \\
 $\Lambda$  & 0.0498  & $-$1.25 &  0.70 & $-$0.09 &  0.50 \\
\hline
&&&& \\
\multicolumn{6}{c}{$\langle \Mapsq \rangle$} \\ \hline\hline
Parameter & $k$ & $\mu$ & $\nu$ & $\eta$ & $\lambda$ \\ \hline
 $\Omegam$  & 0.0073  & $-$0.90 &  0.51 & $-$0.02 &  0.70 \\
  $\Gamma$  & 0.0085  & $-$1.20 &  0.67 & $-$0.05 &  1.03 \\
$\sigma_8$  & 0.0086  & $-$1.03 &  0.58 & $-$0.05 &  1.00 \\
     $z_0$  & 0.0131  & $-$1.11 &  0.62 & $-$0.04 & $-$0.49 \\
$n_{\rm{s}}$  & 0.0184  & $-$1.23 &  0.68 & $-$0.04 &  1.00 \\
   $\beta$  & 0.0172  & $-$1.11 &  0.62 & $-$0.04 &  0.45 \\
 $\Lambda$  & 0.0424  & $-$1.23 &  0.68 & $-$0.06 &  0.24 \\
\hline
\end{tabular}
\end{table}

\begin{figure}
\protect\centerline{
\epsfysize = 3.2truein
\epsfbox[23 150 591 716]
{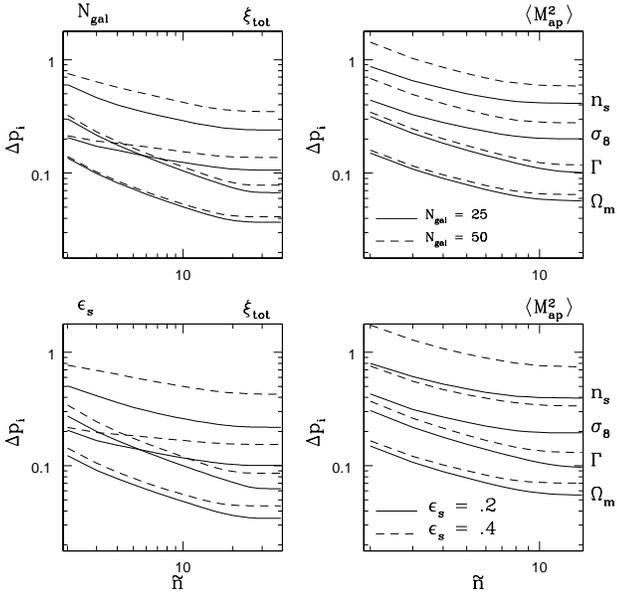}}
\caption{$\Delta p_\alpha = (F_{\alpha\alpha})^{-1/2}$ as a function of the number of
modes $\tilde n$ as in (\ref{tilde-F-cmean}), for different noise
levels.
\emph{Upper panels:} $n_{\rm gal} = 25 \, {\rm arcmin}^{-2}$ (solid
lines), $n_{\rm gal} = 50 \, {\rm arcmin}^{-2}$ (dashed lines). 
\emph{Lower panels:} $\sigma_\varepsilon = 0.2$ (solid lines),
$\sigma_\varepsilon = 0.4$ (dashed lines).
The left panels corresponds to the joint correlation function $\xi_{\rm
tot}$ whereas in the right panels results for $\langle \Mapsq \rangle$
are shown. The plots are done for the parameters $\Omega_{\rm m},
\sigma_8, \Gamma$ and $n_{\rm s}$. The survey geometry is a single
$(30, 100^\prime)$ patch.}
\label{fig:survey_ngal}
\end{figure}

Noise at small angular scales is dominated by Poisson noise coming
from the intrinsic ellipticity dispersion of the galaxies
$\sigma_\varepsilon$, and the finite number density of galaxies
$n_{\rm gal}$. The effect of both error sources is shown in
Fig.~\ref{fig:survey_ngal}. The information content of each eigenmode
decreases with decreasing $n_{\rm gal}$ and increasing
$\sigma_\varepsilon$. The asymptotic plateau is reached for marginal
smaller mode numbers in the case of higher noise level. The resulting
decrease in accuracy with which a given parameter is determined can be
inferred from the asymptotic values.

\subsubsection{Binning}
\label{sec:bin}

\begin{figure}
\protect\centerline{
\epsfysize = 3.2truein
\epsfbox[23 150 591 716]
{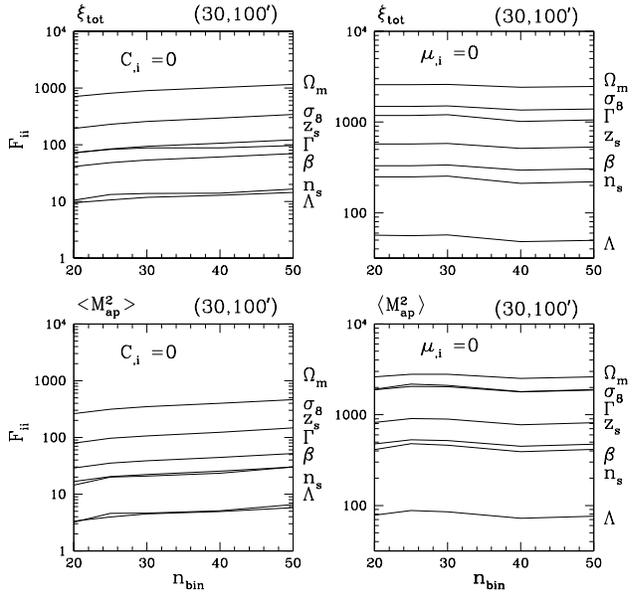}}
 \caption{Effect of the bin number $n$ on the Fisher matrix in the constant
   mean case for various estimators as indicated in the panels.}
\label{fig:bin-size}
\end{figure}

We vary the number of angular bins $n$ of the two-point correlation
functions between 20 and 50, in order to quantify the effect on our
results. For the constant mean case, the power-law decrease of the
Fisher matrix diagonal are unaffected by the binning. 
We found the height of the plateau to weakly depend on the bin
number, as $n^{-\eta}$ where $\eta$ is different for different
statistics and parameters, and ranges between 0.18 and 0.3.

Note that the compression factor is somehow arbitrary since the number
of bins can be chosen freely. However, within the (realistic) range
that we tested, it is close to two. If we further increase the number
of bins, the compression will of course increase; no new information
will be added since the correlation between adjacent bins will
approach unity for very small separations. The weak dependence of the
plateau on $n$ shows that we have not yet reached that saturation
regime. If one considers the observed galaxy ellipticities as input
data vector, the compression factor is of course much larger.

The Fisher matrix in case of constant covariance is only very weakly
dependent on the bin number, as shown in the previous section (see
Table \ref{tab:mvb}).  The diagonal elements of the Fisher matrix for
both cases (constant mean/covariance) are shown in
Fig.~\ref{fig:bin-size}.  In both the cases, for constant covariance
and constant mean, the amplitude of the window functions increases
with the number of bins. The shape is not affected by the binning.

\section{Discussion}

We study the Karhunen-Lo\`eve generalised eigenmode problem in the
context of weak lensing surveys. Four combinations of second-order
shear statistics estimated at various angular scales are used as input
data vectors -- these are the 2PCFs $\xi_{\pm}$, their combination
$\xi_{\rm tot}$ and the aperture mass dispersion $\langle \Mapsq
\rangle$. Different realistic survey geometries are considered to
study the impact of noise and finite sky coverage for these
estimators. We examine the dependence of both the mean and the
covariance of these estimators on various cosmological parameters and
the source galaxy redshift distributions. A Fisher matrix analysis is
presented to determine errors associated with the estimation of
various cosmological parameters and their cross-correlation with
survey parameters.

We consider two different scenarios. In one case, the mean of
second-order shear estimators is used to constrain the cosmological
parameters while the covariance is constant and independent of
cosmology. The second case uses the covariance of these estimators to
constrain cosmological parameters assuming a constant mean. We study
the information content of KL eigenmodes in both cases for various
parameters. For the constant mean case, there are several eigenmodes
among which the information is distributed; however, there is only one
eigenmode associated with the constant covariance case which contains
the complete information. The resulting error bars are
anti-proportional to the square root of the survey area for the
constant covariance case, and independent of the survey area if the
mean is constant. Thus, for reasonable sky coverage (more than about
0.4 square degrees) the first case dominates over the second and the
bulk part of the cosmological information is collected in only one mode.

From the results of our KL eigenmode analysis we find that, starting
from $n$ binned values of second-order shear statistics, a compression
factor of almost two can be achieved in most cases without information
loss, if the cosmological parameters are determined using the
covariance. KL analysis provides rank-ordered, uncorrelated eigenmodes
and corresponding window functions in the case of single parameter
estimation. These sets of window functions provide cleanest measures
of the projected power spectrum for a specific survey strategy. The
first eigenmodes contain most of the information that one can extract
and directly use for maximum-likelihood studies to constrain
cosmological parameters. Typically, the error bars on individual
parameters using the covariance are smallest when the joint correlation
function $\xi_{\rm tot}$ is used. $\xi_+$ gives better constraints
than $\xi_-$ which is in turn better than the aperture mass statistics
$\langle \Mapsq \rangle$.

We provide both an independent study of different parameters as well
as a joint analysis of all parameters for each of these two cases
(constant mean and constant covariance). The parameters which are
near-degenerate have similar KL eigenfunctions. Among the five
cosmological parameters we have considered, ($\Omegam$, $\sigma_8$)
and ($\Gamma$, $n_{\rm s}$) show similar levels of degeneracy and have
very similar eigenmodes. Moreover, the parameters characterising the
source galaxy redshift distribution ($z_0$ and $\beta$, see
(\ref{pzdz})) are degenerate to a high level with $\Omegam$ and
$\sigma_8$.

It may be useful to note that additional constraints may be put by
including higher-order statistics which will be left for further
studies. Moreover, since the number of observables increases with
increasing order (i.e.\ the shear three-point correlation function has
eight independent components and depend on triangle configurations,
Schneider \& Lombardi 2003), data compression can be useful and even
necessary.

The eigenmodes associated with $\Omegam$ and $\sigma_8$ mainly focus
on larger angular scales where cosmological information is not
contaminated by noise on small scales. On the other hand, $\Gamma$ and
$n_{\rm s}$ measure the shape of the projected matter power spectrum
which is easier to determine when also information on smaller angular
scales is available. Therefore the eigenmode windows associated with
these two parameters tend to take more contributions from smaller scales.
In contrast, the eigenmodes associated with the constant covariance
case take contributions from virtually all angular scales probed by
the survey and reaches a maxima roughly at medium angular scales where
the signal dominates over both shot noise and scatter due to finite
sky coverage.

Higher order KL modes do not carry any useful information and
approximately half of the modes contain all of the information content
of the Fisher matrix for a constant mean. However, the off-diagonal
terms of the Fisher matrix which encode the cross-correlation among
various parameters are also a function of the number of modes we
include to reconstruct the Fisher matrix. We found that these terms
keep evolving even after the diagonal terms have already reached a
saturation limit. For very small numbers of modes the reconstructed
Fisher matrix becomes singular, since the information sampled in only
the few eigenmodes is too little in order to put
constraints on more than one parameter, thus different cosmological
parameters become completely degenerate.

The joint analysis of all seven parameters being estimated from the
mean only (constant covariance) shows that only two eigenmodes (out of
seven) are needed to constrain the parameters and to lift their
near-degeneracies as far as possible. 

In our analysis involving KL eigenmodes we have not included and
modelled any systematic measurement errors. We have assumed that the
errors are dominated by Poisson noise and intrinsic ellipticity noise
at small scales and cosmic variance at larger angular scales. The
covariance of the second-order shear statistics include all these
noise sources, which are exact if the shear is a Gaussian
field. Non-Gaussianity leads to an under-estimation of the noise on
angular scales between 1 and 10 arc minutes.

Although the exact implementation of the survey geometry makes our
approach more realistic, we still ignore B-mode contamination due to
systematic residuals. This is mainly due to the absence of a specific
model for the B-mode power spectrum in the case of systematic
measurement errors. Such issues will have to be included in future
studies. Although effects like intrinsic galaxy alignment and source
clustering, which also cause a B-mode, can be and have been modelled
(Crittenden et al.~2001; Croft \& Metzler 2001; Heymans et al.~2004;
Schneider et al.~2002), their contribution to the shear signal is
secondary.

Incorporation of (photometric) redshift information will in general
improve the accuracy with which cosmological parameters can be
recovered. However, we have not divided the source galaxies into
redshift bins, because of the corresponding complication in the
calculation of the covariance. However, this is possible in principle,
e.g.\ using correlated Gaussian fields, see e.g.\ Simon, King \&
Schneider 2004. The inclusion of non-Gaussian terms in the Fisher
matrix analysis (Taylor \& Watts 2001) will invariably help us to
break some of the parameter degeneracies which we have studied
here. However, the construction of covariance matrices associated with
various third-order estimators tend to be very cumbersome. Although
certain analytical calculations were made recently under simple
assumptions (Munshi \& Coles 2003; Munshi \& Valageas 2005), more
detailed analyses which incorporate realistic survey geometries remain
to be done.

We incorporated the non-linear model of the power spectrum according
to Peacock \& Dodds 1996. In a forthcoming work, an improved version
based on the halo model, e.g.\ halofit (Smith et al.~2003) will
be used. In addition, baryonic physics, non-zero neutrino masses and a
varying dark energy equation-of-state parameter will be taken into
account.

A simple-minded signal-to-noise eigenmode analysis (an interesting
special case of KL eigenmode analysis, see TTH) is difficult to
perform for the case of the covariance matrices we have
considered. This is because of the difficulty in separating the signal
and noise contributions in the covariance matrices. Where as the $\Mat
V$ terms (cosmic variance) are pure signal and the $\Mat D$ terms
(shot noise) is pure noise, it is difficult to separate the mixed
terms $\Mat M$ with such clarity. Increasing the noise contribution
either by increasing the intrinsic ellipticity dispersion of galaxies
$\sigma_\ve$ or by decreasing the number density of source galaxies
$n_{\rm gal}$ results in a decrease in the information content of each
of the KL eigenmodes. In addition the number of eigenmodes which
contain useful information reduces, too.

\section*{acknowledgments}
MK was supported by the Deutsche Forschungsgesellschaft (DFG) under
the project SCHN 342/3-1. MK thanks Peter Schneider and Peter Watts
for helpful discussions and useful comments on the manuscript. DM was
supported by PPARC of grant RG28936. It is a pleasure for DM to
acknowledge many fruitful discussions with members of Cambridge
Leverhulme Quantitative Cosmology Group as well as member of Cambridge
Planck Analysis Center. DM also benefitted from useful discussion with
Patrick Valageas, Alan Heavens and Lindsay King. We wish to thank the
anonymous referee for useful suggestions and comments which led to
a substantial improvement of the paper.

\end{document}